# Asymmetric Transmission and Isolation in Nonlinear Devices: Why They Are Different

David E. Fernandes and Mário G. Silveirinha, *Fellow*, IEEE

*Abstract* — Here, we highlight the fundamental differences between nonlinear two-port devices with strongly asymmetric transmission responses and "isolators". We use a mushroom-structure loaded with nonlinear elements as a guiding example. The mushroom-metamaterial can be operated in a regime where it behaves as a nearly ideal electromagnetic diode, such that the individual excitations of the two ports lead to strongly asymmetric responses. We point out the limitations of using this type of nonlinear devices as microwave isolators. In particular, we underline the crucial importance of material loss to attain the isolator functionality and that in the lossless regime nonreciprocal devices may be time-reversal invariant.

## I. INTRODUCTION

RECIPROCITY is a feature of electromagnetic systems resulting from linearity and time-reversal symmetry [1]. This fundamental property has far-reaching consequences, e.g., it implies that reciprocal systems are inherently bi-directional. Hence, microwave devices such as isolators, gyrators and circulators which are commonly used in telecommunications and information processing networks, must rely on nonreciprocal elements. The most typical way to break the Lorentz reciprocity is through the magneto-optic Faraday effect in magnetic materials [2]-[5]. Crucially, the need for a magnetic field bias hinders the use of this solution in highly-integrated devices. There are, however, other alternatives to obtain a nonreciprocal response, namely by exploiting nonlinear effects [6]-[9], with a mechanical motion [10]-[11], with materials with a time-modulated response [12]-[13], or with active elements [14]-[16].

The mushroom structure has attracted a lot of attention due to its applications in the microwave and millimeter wave frequency bands [17]. The mushroom metamaterial consists of a standard wire medium [18]-[20] loaded with metallic patches [19],[21]. These metamaterials have been widely used to enhance the radiation properties of low-profile antennas [17],[22], to obtain negative refraction and near-field imaging [23],[24], in the realization of absorbers for high-power signals [25], just to name a few applications. Furthermore, we showed in [26] that an asymmetric mushroom metamaterial loaded with a nonlinear element may exhibit a strong bistable response and may be used as an electromagnetic switch.

Motivated by these findings, here we demonstrate that the nonreciprocal response of the mushroom-metamaterial can be exploited to design an "electromagnetic diode", understood as a device that for the same intensity of the excitation field is either nearly fully reflecting or nearly fully transparent, depending on the excitation port. The analysis is carried out using the effective medium model developed in [26]-[27] and full wave simulations. We highlight the limitations of this type of nonlinear devices by showing that under a simultaneous excitation they do not behave as microwave isolators. Furthermore, we underscore that the "electromagnetic diode" operation is fully compatible with a design with no material loss, whereas the isolator functionality is not. In particular, it is proven that lossless nonlinear devices may be time-reversal invariant and hence are intrinsically bi-bidirectional.

## II. ASYMMETRIC TRANSMISSION VS. ISOLATION

We consider an asymmetric mushroom metamaterial slab excited by plane waves propagating along the $+z$ and $-z$ directions, i.e., normal to the interfaces, as depicted in Fig. 1. The incident electric field is polarized along the $x$-direction. The wires are modeled as perfect electric conductors (PEC) and are assumed to stand in air. At the top interface the wires are connected to the metallic patches through an ideal short-circuit, whereas at the bottom interface they are connected through a nonlinear load. The wires are misaligned with respect to the geometrical center of the patches [22],[27].

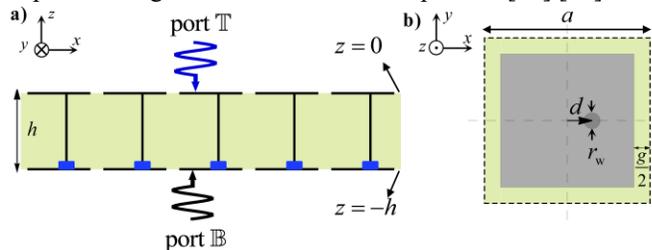

Fig. 1. Geometry of the two-sided mushroom slab a) Side view: The wires are embedded in a dielectric (air) with thickness $h$ and are connected to metallic patches through lumped nonlinear capacitors at the bottom interface, represented in the figure as blue rectangles, and through ideal short-circuits at the top interface. b) Top view of the unit cell: the wires are arranged in a square lattice with period $a=1$ cm and have radius $r_w=0.025a$; they are displaced by a distance $d=a/4$ with respect to the center of the patches. The separation between adjacent patches is $g=0.05a$.

Similar to our previous work [26], for convenience we model the nonlinear load as a parallel plate-type capacitor

This work was funded by Fundação para a Ciência e a Tecnologia (FCT) under projects PTDC/EEI-TEL/4543/2014 and UID/EEA/50008/2013. D. E. Fernandes acknowledges support by FCT, POCH and the cofinancing of Fundo Social Europeu under the fellowship SFRH/BPD/116525/2016.

D. E. Fernandes is with the Instituto de Telecomunicações, Departamento de Engenharia Electrotécnica, Universidade de Coimbra - Pólo II, 3030-290 Coimbra, Portugal (e-mail: dfernandes@co.it.pt). M. G. Silveirinha is with the University of Lisbon and Instituto de Telecomunicações, Avenida Rovisco Pais, 1, 1049-001 Lisboa, Portugal (e-mail: mario.silveirinha@co.it.pt)

filled with a nonlinear Kerr-type dielectric with third-order electric susceptibility $\chi^{(3)} = 0.9 \times 10^{-9} \, \text{m}^2 \text{V}^{-2}$. It was shown in [26] that this lumped element may provide a nonlinearity strength comparable to that of a commercially available varactor. The capacitance of the nonlinear element is $C_L = \varepsilon_0 \varepsilon_{\text{cap}} l_{\text{cap}}^2 / t_{\text{cap}}$, where $l_{\text{cap}}^2$ corresponds to the capacitor cross-sectional area, $t_{\text{cap}}$ to its thickness and $\varepsilon_{\text{cap}}$ is the permittivity of the Kerr-material, which may be written as a function of the voltage across the capacitor [26]. In the linear regime, the permittivity is taken equal to $\varepsilon_{\text{cap}}^0 = 2$. The structural parameters of the capacitor are $l_{\text{cap}} = 3g$ and $t_{\text{cap}} = 2g$. For further details about the geometry of the nonlinear element the reader is referred to [26].

*A. The linear regime*

To begin with, we study the linear response of the asymmetric mushroom-structure when it is illuminated separately from the top and bottom sides (port $\mathbb{T}$ and port $\mathbb{B}$, respectively, as shown in Fig. 1) for a time variation of the type $e^{j\omega t}$, with $\omega$ the oscillation frequency. To have a strong nonlinear response, the voltage across the capacitor should be highly sensitive to small variations of the excitation field [26]. For the considered structural parameters, the strongest resonance of the metamaterial occurs for a Fano resonance near 14.6GHz [26]. Therefore we focus our attention in a spectral range nearby this frequency. Figures 2a) and 2b) depict the transmission and reflection coefficients as function of frequency for an individual excitation of ports $\mathbb{T}$ (dashed lines) and $\mathbb{B}$ (solid lines). Due to the reciprocity of the involved materials, in the linear regime the transmission coefficient is independent of the excitation port (both amplitude and phase). Because all the materials are taken as lossless in the simulation, the amplitude of the reflection coefficient is also independent of the excitation port. In contrast, due to the structural-asymmetry of the mushroom structure (non-centered wires and lumped load near the bottom plate) the phase of the reflection coefficient depends slightly on the direction of arrival of the incoming wave.

Importantly, despite the transmission level being identical for both excitations, the fields inside the slab can be rather different due to the structural asymmetry [7]. Indeed, the electric field at the resonance can be substantially larger for a top excitation as compared to a bottom excitation (see Fig. 2c). Later we will consider that the capacitor element has a nonlinear response. In the linear regime, the fields inside the capacitor are proportional to the feeding current [27]. Therefore, a difference between the currents induced by each excitation will result in an asymmetric nonlinear response. In Fig. 2d we depict the profile of the current $I$. The homogenization results are validated using the full-wave electromagnetic simulator CST-MWS [28]. Similar to the electric field inside the slab, the load current $I|_{z=-h}$ is higher for a $\mathbb{T}$-port excitation. Therefore, it is expected that for the same amplitude of the incident field, a top excitation will lead to more pronounced nonlinear effects.

It is convenient to introduce the transfer functions $V_L / E^{\text{inc}, \mathbb{T}} \equiv 1/F^{0,\mathbb{T}}$ and $V_L / E^{\text{inc}, \mathbb{B}} \equiv 1/F^{0,\mathbb{B}}$, which represent the ratio between the voltage induced at the lumped load ($V_L$) and the incident field complex amplitude for top and bottom excitations. Evidently, the transfer functions depend on frequency and on the load value $F^{0,i} = F^{0,i}(Z_L)$, $i = \mathbb{T}, \mathbb{B}$, where $Z_L$ is the lumped element impedance. The two transfer functions can be numerically determined using the effective medium model described in [26]. For a simultaneous excitation of the two ports, one has $V_L = E^{\text{inc},\mathbb{T}}/F^{0,\mathbb{T}} + E^{\text{inc},\mathbb{B}}/F^{0,\mathbb{B}}$ by linearity.

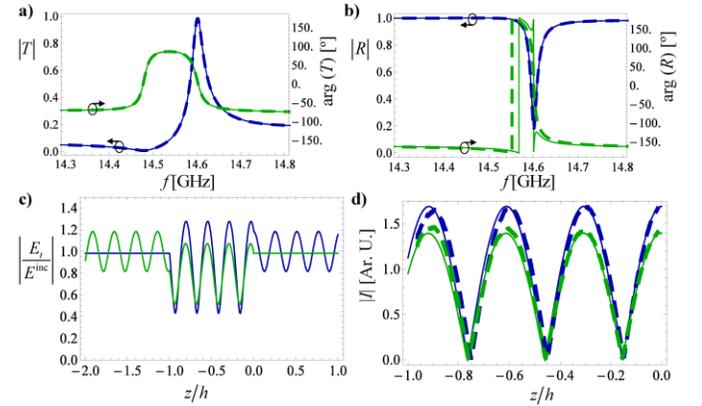

Fig. 2. Amplitude and phase of the a) transmission and b) reflection coefficients of the mushroom metamaterial in the linear regime The solid (dashed) curves correspond to an excitation of the $\mathbb{B}$ ($\mathbb{T}$) port, respectively. c) and d) Profiles of the *x*-component of the electric field and of the current along the wires, respectively, at the resonant frequency 14.60GHz. The green/blue curves are for a bottom/top excitation. In d) the solid curves correspond to the effective medium model results and the dashed curves to the full-wave simulations.

*B. The nonlinear regime*

A simple generalization of the theory of [26] shows that because the nonlinear response is concentrated in the lumped element, the relation $V_L = E^{\text{inc},\mathbb{T}}/F^{0,\mathbb{T}} + E^{\text{inc},\mathbb{B}}/F^{0,\mathbb{B}}$ remains valid in the nonlinear regime. Note that we consider only the nonlinear effects on the first harmonic, i.e., the nonlinear self-action [29]. In the nonlinear regime $Z_L$ is some known function of $|V_L|$ [26]. In particular, the point of operation of the system (determined by $|V_L|$) can be found by solving the nonlinear equation

$$|V_L| = \left| E^{\text{inc},\mathbb{T}}/F^{0,\mathbb{T}}(Z_L) + E^{\text{inc},\mathbb{B}}/F^{0,\mathbb{B}}(Z_L) \right|, \quad (1)$$

with respect to $|V_L|$ with $Z_L = Z_L(|V_L|)$. Once $|V_L|$, and thereby also the impedance $Z_L$ are found, all the fields can be determined from the solution of the linear problem [26].

Similar to Ref. [26], in the following we fix the frequency at 14.56GHz. This frequency corresponds to a point of a strong reflection (see Fig. 2a), and hence in the linear regime (for weak incident fields) the metamaterial slab will block an incident wave.

Figure 3a depicts the transfer functions $F^{0,\mathbb{T}}$ and $F^{0,\mathbb{B}}$. As seen, $|F^{0,\mathbb{T}}|$ is smaller than $|F^{0,\mathbb{B}}|$ for the same load. This property implies that the voltage induced at the load is larger for top incidence, which is consistent with the results of Fig. 2d for the current along the wires. Since the load is purely reactive, the transfer functions approach infinity in the limit $X_L \to 0$.

For an individual excitation of the ports we can write $|E^{\text{inc},i}| = |V_L||F^{0,i}(Z_L)|$ with $i = \mathbb{T}, \mathbb{B}$, and hence the incident field may be regarded as a function of $|V_L|$ [26]. The results of Fig. 3b show that the relation between the incident field and the load voltage is not univocal. This property leads to a bistable response, which is physically manifested in the form of hysteresis loops [26]. This is shown in Figs. 3c-d which depict the scattering parameters determined with the effective medium theory and with full-wave simulations [28], as a function of the amplitude of the excitation field at ports $\mathbb{T}$ and $\mathbb{B}$. The agreement between the two sets of curves confirms the validity of the homogenization method. Each of the two ports can be operated as an electromagnetic switch controlled by the amplitude of the incident field. However and most importantly, the response of the two ports can be strongly asymmetric. In particular, in a scenario where the excitation field decreases from very large values to around $|E^{\text{inc}}| = 500 \text{ V/m}$ (highlighted in Fig. 3c-d as a light brown region) we have $|T_\mathbb{T}| \approx 1$ and $|T_\mathbb{B}| \approx 0.15$ for the top and bottom excitations, respectively. Thus, in this regime the mushroom is nearly transparent/opaque for a top/bottom excitation with the same amplitude, similar to an electronic diode but for microwave signals.

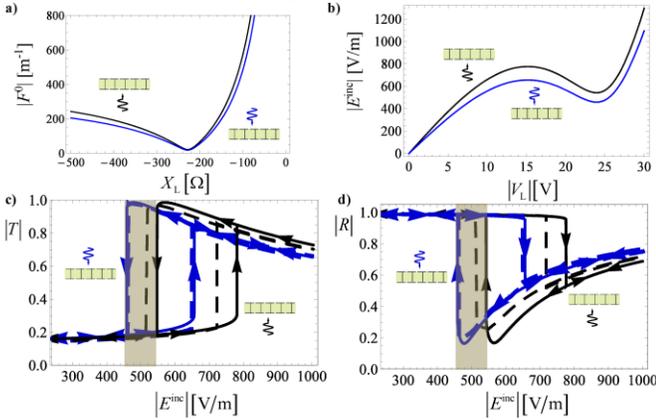

Fig. 3. a) Transfer functions $|F^{0,\mathbb{T}}|$ and $|F^{0,\mathbb{B}}|$ as a function of the reactance of the lumped capacitive element. b) Amplitude of the excitation field as a function of the voltage at the nonlinear capacitor. c) and d) Transmission and reflection coefficients as function of the amplitude of the incident field. The arrows indicate whether the excitation field is increasing or decreasing. The solid lines represent the effective medium model results and the dashed lines the full-wave simulations. In all panels, the blue (black) curves correspond to an excitation at port $\mathbb{T}$ ($\mathbb{B}$). The operating frequency is 14.56GHz.

It is interesting to note that by taking advantage of the bi-stable response of the structure, e.g., by enforcing a suitable biasing so that the port $\mathbb{T}$ is operated in the upper branch of the hysteresis cycle and the port $\mathbb{B}$ in the lower branch, it is possible to have a wide intensity range for $|E^{inc}|$ (in our design, $450 < |E^{inc}| < 750 \, [\text{Vm}^{-1}]$) where $|T_\mathbb{T}|/|T_\mathbb{B}| \gg 1$ and $|T_\mathbb{T}|$ is near unity. Thus, bistability provides a route to surpass some of the limitations of Fano-type nonlinear devices identified in [9]. Furthermore, it was recently shown that by using coupled nonlinear resonances it is possible to have a broadband electromagnetic diode operation fully transmitting from one side and fully reflecting from the other side [30]. However, one should keep in mind that such strategies are only valid for the *individual* excitations of the ports. This property is further highlighted in the next sub-section.

*C. Simultaneous excitations*

The results of subsection II-B may suggest that the proposed lossless device can be used as a microwave isolator of a high-power source placed at the $\mathbb{T}$ port side. Ideally, the level of the signal emitted by the source should lie somewhere in the light brown shaded region of Fig. 3c, which, of course, can be controlled by design.

To model the effect of a possible impedance mismatch at port $\mathbb{B}$, we consider next the simultaneous excitation of the two ports with $E^{\text{inc},\mathbb{B}} = be^{i\phi} E^{\text{inc},\mathbb{T}}$ with $b > 0$. Note that $be^{i\phi}$ represents simply the ratio of the incident fields and is not the reflection coefficient at port $\mathbb{B}$. From Eq. (1) and using $E^{\text{inc},\mathbb{B}} = be^{i\phi} E^{\text{inc},\mathbb{T}}$, one finds that for a simultaneous excitation $V_L = E^{\text{inc},\mathbb{T}}/F^0$ with $1/F^0 = 1/F^{0,\mathbb{T}} + be^{i\phi}/F^{0,\mathbb{B}}$. Thus, we can use the same methods as in subsection II.B to characterize the influence of the $|E^{\text{inc},\mathbb{T}}|$ on the scattered fields.

In order to illustrate the effect of simultaneous excitation it is supposed that $be^{i\phi} = 0.75e^{j0°}$, which may model a moderate impedance mismatch at port $\mathbb{B}$. The incident electric fields at the two-ports are in phase ($\phi = 0°$). In order to place the point of operation of the $\mathbb{T}$ port in the upper-branch of the hysteresis cycle, the structure is excited with a time-varying amplitude modulated signal with $|E^{inc}| = 900 \, \text{V/m}$ for $t < 350\text{ns}$ and $|E^{inc}| = 475 \, \text{V/m}$ when $t > 350\text{ns}$ (green curve in Figs. 4a and 4b).

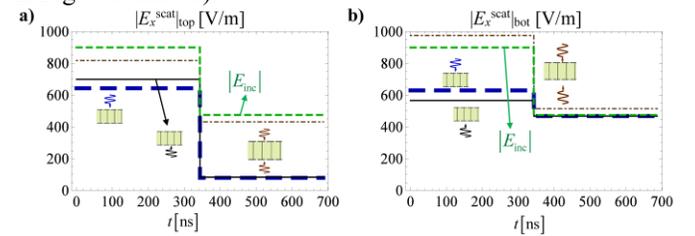

Fig. 4. Envelope of the scattered field at the a) top and b) bottom interfaces as a function of time. Solid black curves: $\mathbb{B}$-port excitation. Thick dashed blue curves: $\mathbb{T}$-port excitation. Dot-dashed brown curves: scenario of simultaneous excitation with $be^{i\phi} = 0.75$. Dashed green curve: incident electric field envelope.

The envelopes of the fields scattered at the two interfaces (amplitude of the wave propagating away from the interface) are represented in Fig. 4 as a function of time for the individual $\mathbb{T}$- and $\mathbb{B}$- port excitations (blue and black curves, respectively). In agreement with Figs. 3c and 3d, for the individual excitations the scattered electric field is large at the bottom interface and rather weak at the top interface. This is the behavior expected from a device that isolates a microwave source in the $\mathbb{T}$-region from the $\mathbb{B}$-region. Importantly, under a simultaneous excitation of the two-ports (brown curves in Fig. 4) the scattered field in the $\mathbb{T}$-region becomes comparable to the incident field. This property is ultimately a consequence of the fact that the nonlinearity of the device spoils the superposition principle. Thereby, the considered device *cannot* be used as an isolator.

*D. The role of material loss and time-reversal invariance*

Another important aspect in the operation of nonlinear devices is the crucial role of material loss in the isolator functionality. It is impossible to realize an isolator without material loss. Indeed, it is unfeasible to isolate a source from an arbitrary load with a nonlinear lossless device. For example, if for a time-harmonic excitation the reflections back to the source could be weak for a *fully reflecting* load (e.g., a short-circuit), this would imply that the energy stored in the nonlinear device / output port would grow steadily in time, which is physically absurd. Moreover, as shown next, lossless nonlinear devices are intrinsically bi-directional.

We consider an arbitrary nonlinear device such that the effects of the nonlinearity are described by some scalar space-dependent permittivity that depends on the intensity of the electric field $\varepsilon = \varepsilon(\mathbf{r}, |\mathbf{E}|)$ and that material loss is negligible. The corresponding source-free nonlinear Maxwell equations in the *time-domain* read (in this subsection $\mathbf{E}, \mathbf{H}$ represent the instantaneous time-domain electromagnetic fields rather than complex amplitudes):

$$\nabla \times \mathbf{H} = \partial_t \mathbf{D}, \qquad \nabla \times \mathbf{E} = -\mu_0 \partial_t \mathbf{H},$$
$$\mathbf{P}(\mathbf{r},t) = \varepsilon_0 \chi(\mathbf{r}, |\mathbf{E}(\mathbf{r},t)|) \mathbf{E}(\mathbf{r},t), \qquad (2)$$

where $\partial_t = \partial/\partial t$, $\mathbf{D} = \varepsilon_0 \mathbf{E} + \mathbf{P}$ is the electric displacement vector, $\mathbf{P}$ is the polarization vector and $\chi$ is the electric susceptibility that determines the nonlinear response of the materials such that $\varepsilon(\mathbf{r}, |\mathbf{E}|) = \varepsilon_0 [1 + \chi(\mathbf{r}, |\mathbf{E}(\mathbf{r},t)|)]$. More generally, one may consider that $\mathbf{P}$ and $\mathbf{E}$ are linked by some differential equation so that material dispersion is taken into account. For example, for a Lorentz-type dispersion one may use $\partial_t^2 \mathbf{P} + \omega_0^2 \mathbf{P} = \varepsilon_0 \omega_0^2 \chi(\mathbf{r}, |\mathbf{E}(\mathbf{r},t)|) \mathbf{E}$, where $\omega_0$ is the resonant frequency. Importantly, in the absence of material loss the considered nonlinear system is *time-reversal symmetric* (even though it is nonreciprocal due to the breach of the linearity condition). Indeed, $\mathbf{P} = \varepsilon_0 \chi(\mathbf{r}, |\mathbf{E}|) \mathbf{E}$ is evidently "even" under a time-reversal operation (the same property holds even in presence of material dispersion). This implies that if some dynamic fields $\mathbf{E}(\mathbf{r},t)$, $\mathbf{P}(\mathbf{r},t)$ and $\mathbf{H}(\mathbf{r},t)$ satisfy the nonlinear time-domain Maxwell equations (2), then the time-reversed fields $\mathbf{E}(\mathbf{r},-t)$, $\mathbf{P}(\mathbf{r},-t)$ and $-\mathbf{H}(\mathbf{r},-t)$ also do.

Consider now a (time-domain) scattering problem in a nonlinear two-port device such that some incident wave in port 1 ($E^{\text{inc}}(t)$) generates reflected ($E^{\text{ref}}(t)$) and transmitted ($E^{\text{tx}}(t)$) waves in ports 1 and 2, respectively. The time-reversal property implies that if the nonlinear device is illuminated with the time-reversed scattered fields (i.e., the time-reversed reflected and transmitted waves) it is possible to return all energy to the source (i.e., to generate the time-reversed incident wave that propagates back to the source). Note that this result is *not* restricted to time-harmonic signals: the time-variation of the input signal can be arbitrary. Furthermore, the effects of frequency mixing are fully taken into account by our theory. In particular, let us suppose that the nonlinear device enables an almost full transmission from port 1 to port 2 such that $E^{\text{tx}}(t)$ is a strong signal and $E^{\text{ref}}(t)$ is a weak signal. Then, if the nonlinear device is *simultaneously* illuminated with the strong signal $E^{\text{tx}}(-t)$ in port 2 and with the weak signal $E^{\text{ref}}(-t)$ in port 1, then all the energy is rerouted to port 1 and generates the pulse $E^{\text{inc}}(-t)$ (propagating in the guide associated with port 1 away from the nonlinear device). Thus, lossless nonlinear devices are inherently bi-directional, independently if they operate in pulsed or continuous wave regimes, and cannot possibly be operated as *robust* optical isolators. Crucially, our result does *not* forbid the ideal electromagnetic diode operation under some *common* excitation conditions (not linked by time-reversal symmetry) [30]. Indeed, for an individual excitation of port 2 (e.g., with only the signal $E^{\text{tx}}(-t)$ as the incident wave) the device can be strongly reflecting. However, its behavior can change dramatically under a simultaneous excitation of the two ports, even if the excitation of the port 1 is rather weak (e.g., with the signal $E^{\text{ref}}(-t)$).

To conclude we note that generic lossy nonlinear devices in continuous-wave (time-harmonic) operation are also constrained by the so-called "dynamic" reciprocity [31]. This property follows from the observation that for a fixed excitation the nonlinear device behaves as a regular dielectric with a non-uniform permittivity profile [31]. Dynamic reciprocity implies that a nonlinear device cannot provide isolation for arbitrary backward-propagating weak signals. Clearly, as shown by our analysis, the absence of material absorption further constraints the response of nonlinear systems.

III. CONCLUSION

In this work we highlighted the fundamental differences between "electromagnetic-diodes" and "isolators". In particular, we theoretically and numerically showed that a lossless mushroom-metamaterial nonlinear device may behave as an almost ideal electromagnetic-diode, but that it does not provide "microwave isolation". Part of the difficulties stem from the inapplicability of the superposition principle and from (quasi-) time-reversal invariance. In addition, the isolation functionality requires material absorption: it is fundamentally impossible to realize an isolator without material loss, both in pulsed and continuous wave regimes.